# Depression Diagnosis and Drug Response Prediction via Recurrent Neural Networks and Transformers Utilizing EEG Signals

Abdolkarim Saeedi[1], Arash Maghsoudi[1,2,3], Fereidoun Nowshiravan Rahatabad[1],

[1] Department of biomedical engineering, Science and research branch, Islamic Azad university, Tehran, Iran

[2] Department of Medicine, Baylor College of Medicine, Houston, TX, United States

[3] Center for Innovations in Quality, Effectiveness, and Safety, Michael E. DeBakey VA Medical Center, Houston, TX.

*Abstract*— *Objective:* **The Early diagnosis and treatment of depression is essential for effective treatment. Depression, while being one of the most common mental illnesses, is still poorly understood in both research and clinical practice. Among different treatments, drug prescription is widely used, however the drug treatment is not effective for many patients. In this work, we propose a method for major depressive disorder (MDD) diagnosis as well as a method for predicting the drug response in patient with MDD using EEG signals.** *Method:* **We employ transformers, which are modified recursive neural networks with novel architecture to evaluate the time dependency of time series effectively. We also compare the model to the well-known deep learning schemes such as CNN, LSTM and CNN-LSTM.** *Results:* **The transformer achieves an average recall of 99.41% and accuracy of 97.14% for classifying normal and MDD subjects. Furthermore, the transformer also performed well in classifying responders and non-responders to the drug, resulting in 97.01% accuracy and 97.76% Recall.** *Conclusion:* **Outperforming other methods on a similar number of parameters, the suggested technique, as a screening tool, seems to have the potential to assist health care professionals in assessing MDD patients for early diagnosis and treatment.** *Significance*: **Analyzing EEG signal analysis using transformers, which have replaced the recursive models as a new structure to examine the time dependence of time series, is the main novelty of this research.**

## I. INTRODUCTION

MAJOR Depressive disorder is a debilitating medical condition that negatively affects a person's feelings, thoughts, and actions. It also causes constant discomfort and loss of daily activities. It affects more than 350 million people worldwide, in addition, the World Health Organization (WHO) predicts that it will become the second leading cause of disability by 2030 [1]. Among depressive disorders, major depression has the highest incidence. Major depression, also known as MDD, is one of the most challenging disorders facing humanity. The rate of people who suffer from it has steadily increased in the recent decades. This prompted the World Health Organization to launch a "Let's Talk About Depression" program in 2017 to educate people about the disorder. Furthermore, depression has a profound effect on the health of the patient, the quality of individual and family life, activities of daily living, as well as on health care providers, payers and employers. People with depression can develop multiple illnesses at the same time, which can have serious side effects and increase costs. The economic burden of this disease is significant. Especially in primary care, most patients with depression are not diagnosed and treated. Early diagnosis, intervention and appropriate treatment can improve, prevent recurrence and reduce the emotional and financial burden of the disease. One of the tools to achieve this, is analyzing the Electroencephalography (EEG) signals which record electrical activity and brain waves using electrodes placed on the scalp. Measuring the electrical activity of the brain is useful because it shows how many different nerve cells in the brain network communicate with each other through electrical impulses. This signal has some unique characteristics which are, high frequency resolution and being cheap and portable compared to MRI for instance. In the meantime, using such a non-invasive method, along with automated deep learning methods and starting treatment courses in a timely manner seems desirable. Figure 1 shows the difference between normal and depressed EEG signals:

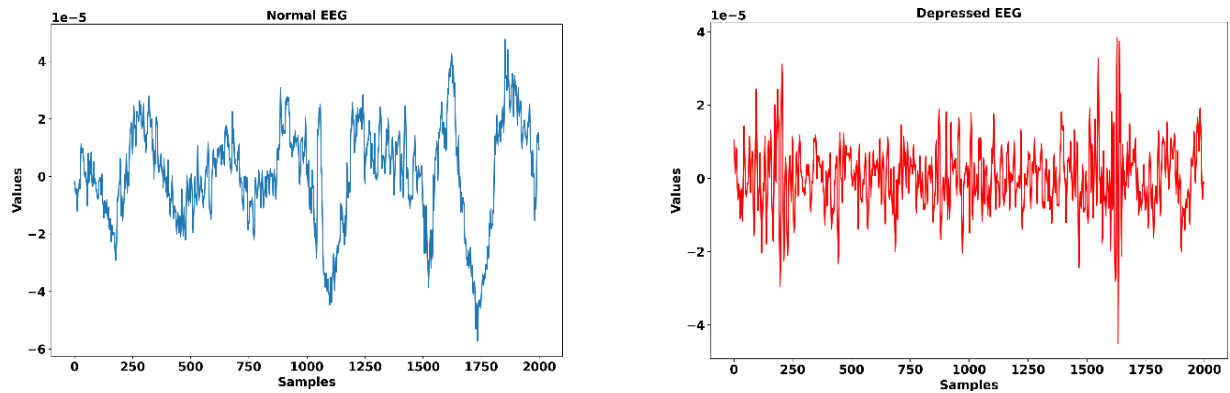

Fig. 1. Examples of EEG from normal and depressed participants

## A. Previous Work

Bairy et al. [2] used discrete cosine transforms to analyze 2,100 recorded samples from 60 individuals (30 healthy, 30 depressed). The authors extracted nonlinear properties such as sample entropy, De-trended Fluctuation Analysis, correlation dimension, fractal dimension and *largest Lyapunov exponent* (LLE). The highest accuracy of 93.8% was assigned to the support vector machine with RBF kernel. Mohammad et al. [3] recorded 3-minute signals from 53 patients with chronic depression and 43 age-matched healthy individuals, using discriminatory linear analysis and then genetic algorithm to reduce feature and model sizes. The objective was select the most useful feature space. Most success was achieved using the decision tree classification algorithm, which was about 80% accurate. Arns et al. [4] conducted an international study to predict optimal treatment for depression. The dataset included 1,088 participants with chronic depression and 336 participants in the control group, each signal had two minutes of recording from open and closed eye states. This study showed that frontal alpha asymmetry and the power of elliptical and frontal biometric alpha are not good for distinguishing between depressed and healthy individuals. The study suggested that alpha asymmetry was related to citalopram and sertraline. Acharya et al. [5] studied 30 participants, each of whom consisted of signals with 2000 samples (7.8 seconds). This study provided a diagnostic indicator for depression to automatically distinguish between normal and depressive individuals. They extracted nonlinear features from EEG signals consisting of fractal dimensions, LLE, sample entropy, oscillation analysis, Hurst exponent, higher order spectra, and recurrence quantification analysis. The best features were selected using t-test and the support vector machine classification achieved 98% accuracy. Bachmann et al. [6] introduced a simple method for the diagnosis of depression depending on single channel EEG based on spectral asymmetry index (SASI) and analysis of nonlinear fluctuations in 17 depressed patients and 17 controls, using the first 5 minutes of each recording. The SASI attribute finds an estimated difference between the power in the two selected high-frequency and low-frequency sub-bands, while the middle band, which is alpha here, is not inserted. Using a combination of these characteristics and linear discriminant analysis (LDA) as a classifier, an accuracy of 91.2% was achieved. Mumtaz et al. [7] performed an experiment on 33 patients with chronic depression and 30 normal participants and utilized linear features such as bandwidth at different frequencies and alpha hemisphere asymmetry of five EEG signals. Their study used the *Receiver* Operating Characteristics (ROC) to rank the most important features affecting classification accuracy, the highest performance obtained was with the Naive Bayes classifier reaching 98.4%. Liao et al. [8], used a proposed new Kernel Eigen-Filter-Bank Common Spatial Patterns (KEFB-CSP) method, the dataset consisted of 12 depressed individuals with 12 normal individuals, the methodology included the following steps: Initially, the signal is decomposed to the desired sub bands. After the spatial domain transformation, a core principal component analysis is performed to convert the current data to its final form (KEFB-CSP), this study yielded the results with a signal of a duration of only 6 seconds, reaching 80% accuracy. Cai et al. [9] collected a database of 86 depressed patients and 92 healthy controls. In this study, a multimedia EEG was used, which included recording in three modes: neutral, negative, and positive sound stimuli. In another study, Cai et al. [10] collected a data set that included 92 patients with chronic depression and 121 normal individuals. The electroencephalogram was 5 minutes long and was recorded from three electrodes. Two hundred and seventy features are reduced by redundancy-maximum correlation method and the selected space is fed to classification algorithms such as decision tree, support vector machine, nearest neighbor method and artificial neural. The K-nearest neighbors technique method gained the highest accuracy 86.98%. Acharya et al. [11] analyzed 5 minutes of EEG signals from 15 normal patients and 15 depressed patients to provide an automatic diagnosis of depression based on a one-dimensional convolutional neural network. The 13-layer CNN network introduced consists of five standard layers, five pooling layers, and three fully connected layers. They also showed that right hemisphere EEG signals are more efficient for deep learning classification with an accuracy of 96% compared to 93.5% of right hemisphere. Wu et al. [12] used positive images as stimuli that extracted relative band characteristics from the electroencephalogram. The authors used a new classification method, the CK support vector machine, and compared it with other machine learning techniques. Participants in this study included 55 people, 24 depressed and 31 healthy. This study achieved 83.64% accuracy. Azizi et al. [13] examined the six-second signals from 40 subjects (20 in each group) and presented a number of new differences based on geometric methods: center-to-center distance, shortest center-to-45-degree line distance, and the in-center radius. They found that the features of the right hemisphere are more appropriate between the two groups. Saeedi et al. [14] examined the eight-second data of 63 individuals (33 depressed and 30 healthy). Frequency characteristics were extracted from different electroencephalogram bands. Discrete wavelet transform and nonlinear properties were also used to better display the data. This study presented a novel method for enhancing the nearest neighbor algorithm utilizing the genetic algorithm and reached 99.44% accuracy. Z. Wan et al. [15] developed a deep learning architecture calling it HybridEEGNet, which involved a combination of two networks for simultaneous and regional extraction. This dataset consisted of 35 participants (12 patients without a prescription, 12 drugs, 11 normal) and the model acquired an accuracy of 79.08%. Fitzgerald et al. [16] reviewed many studies investigating the effect of gamma waves on feature extraction from electroencephalogram signals and their power of detection. They also concluded that a further usage of this rhythm should be considered in future studies. Ay et al. [17] used 30 individuals (15 normal, 45 depressed patients) and raw signals for their experiments. 2000 samples consisting of 7.8 s with a sampling frequency of 256 Hz to prepare a deep learning model using the convolutional neural network first to extract features and reduce variance, then feed the convolutional neural network maps directly to an LSTM cell and use They used fully connected layers for classification. The CNN-LSTM model achieved 99.12% and 97.06% for the right and left hemispheres, respectively. Sharma et al. [18] developed a new band-stop energy (SBE) method using wavelet transform to decompose signals into sub-bands. The study included 2130 participants and used the support vector machine for the classification purpose, the accuracy was 34.79%. Saeedi et al. [19] evaluated the performance of one-dimensional and two-dimensional CNNs and CNN-LSTMs on Generalized Partial Directed Coherence (GPDC) and Direct directed transfer function (dDTF) brain connectivity matrices and reached 99.24% accuracy on the same dataset used in [7]. Mahato et al. [20] recorded a five-minute study of 34 patients with chronic depression and 30 healthy volunteers. This study extracted linear band characteristics from alpha, alpha 1, alpha 2, beta, delta, theta power and theta asymmetry values. They used multi-cluster feature selection to select effective features. The highest accuracy (88.33%) was obtained using a combination of alpha 2 and theta asymmetry while the support vector machine was distinctive. Čukić et al. [21] evaluated the effectiveness of two widely used measures, Higuchi's fractal dimension and Sample Entropy. The data set included 23 depressed patients and 20 healthy individuals. The study compared logistic regression, support vector machine, decision tree, random forests and simple bayes. They concluded that nonlinear methods are practical for discriminating chronic depression. It is also stated that linear features can be reasonably accurate. This research obtained accuracies in the range of 90.24% to 97.56% using a combination of the aforementioned algorithms. Analyzing EEG signal analysis using transformers, which have replaced the recursive models as a new structure and examine the time dependence of time series, is the main novelty of this research. Other objectives that were acquire in this study are: 1- Ability to classify healthy and depressed EEG signals with appropriate accuracy. 2- Comparison of transformers with different types of recurrent neural networks. 3- Diagnosis of the drug-responsive group in the depressed group. Figure 2 shows the outline of the process:

## II. Materials

### A. Participants

The data set used in this article has been prepared by Mumtaz et al. [7] which is open to the public. The recordings included 34 depressed patients, 17 women aged 27 to 53 years and 30 normal people (9 women) aged 22 to 53 years.

### B. Data acquisition

EEG data were recorded for 5 minutes in the closed eye state. The recordings were obtained from 19 cap electrodes mounted on the scalp according to the 10-20 classification of international standard electrodes. The sampling frequency was 250 Hz. A gap filter was used to reject 50Hz power line noise. All EEG signals with a cut-off frequency at 0.5 Hz and 70 Hz were passed through the band filter.

### C. Preprocessing

Two thousand samples of the 57 participants with 20 channels for each person will be converted into baseline datasets. The process includes selecting a range of 2000 samples to extract from the signals resulting in a (57, 20, 2000) array, after that each channel is concatenated into the first dimension treating them as individual samples, which results in 1102 records of two thousand examples with a shape of (1102, 2000). To diagnose the responder group in depressed people, 30 records is converted into 570 data samples with the same approach.

## III. Methods

### A. Convolutional neural networks (CNN)

CNN is a neural network that is commonly utilized in image processing, machine vision and classification tasks. It is one the most advanced deep learning methodology, consisting of multiple stacked convolutional layers. This network usually consists of convolutional layers, pooling layers and fully connected layers [24].

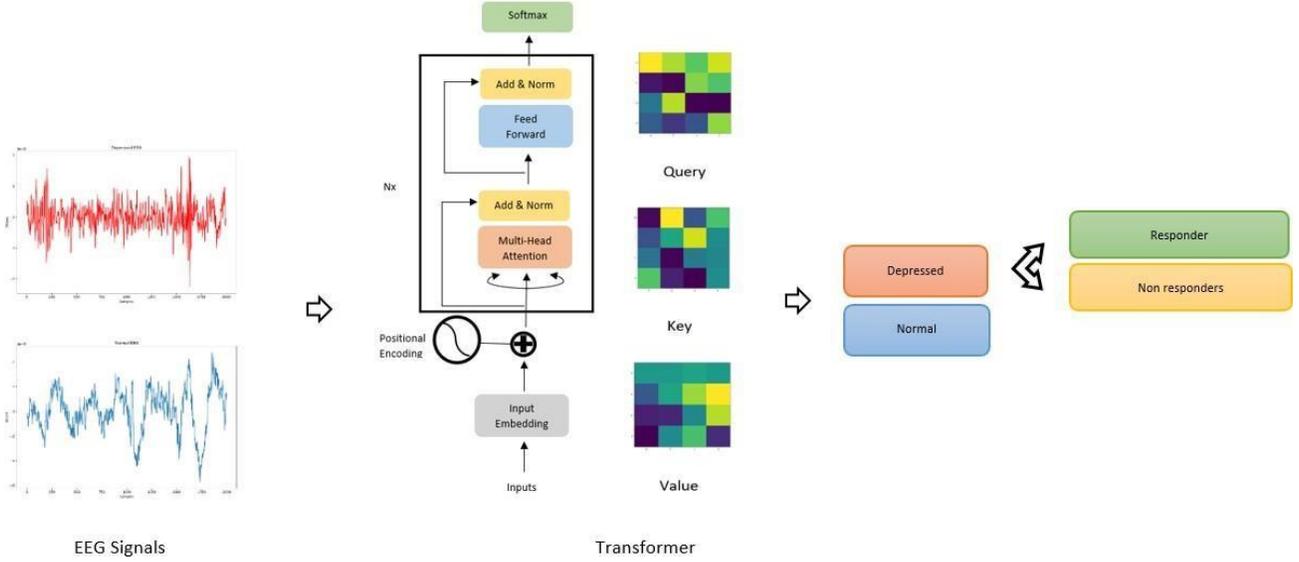

Fig. 2. Block diagram of the proposed method which summarizes the process of the entire work

### *Convolutional layers*

In this layer, the data is subjected to a linear transformation using specific filters. Filters are used for extracting essential patterns from prototypes. The following is a definition of this operation:

$$x_{ij}^l = \sum_{a=0}^{k-1}\sum_{b=0}^{k-1} w_{ab}\, y^{l-1}\,(i+a)(j+b) \qquad (1)$$

$$y^i = \sigma(x^i) \qquad (2)$$

Where the filter is a $K \times K$ matrix, each element of which corresponds to a weight in $w_{ab}$. $x_{ij}^l$ denotes a single output element from the current layer $l$ and is calculated by sliding the filter over the previous layer's output $y^{l-1}$. Following the convolution procedure, a sigmoid activation function $\sigma$ is applied to $x_{ij}^l$, introducing non-linearity, as seen in equation (10). To produce the most distinct features, filter coefficients are modified throughout the training phase depending on error back propagation. As a result, depending on the input parameters, filters can have a varying count, length, stride, and padding size.

### *Pooling layers*

Pooling, like convolutional layers, is important in CNNs. A pooling layer has a fixed kernel size and walks through the feature map produced by convolutional layers, reducing dimensionality by picking only a subset of the available elements. The input size in a convolutional layer is decreased from $N \times N$ to $N - m + 1 \times N - m + 1$, however in a pooling operation it is reduced to $N/K \times N/K$ with a pool size of $K \times K$ since only one element is chosen in each iteration. A mean, max, or summation operation can be selected depending on the layer type. This process speeds up calculations while also making the network less exposed to every single element in the feature map. It also helps to minimize overfitting by decreasing parameters and making the network less vulnerable to every single element in the feature map.

### *Fully-connected layers*

The feature map is eventually flattened to a single column vector and fed into fully-connected layers. They enable non-linear feature summation to be employed to maximize

discrimination. The entire procedure is as follows:

$$v_j(n) = \sum_{i=0}^{m} w_{ji}(n) y_i(n) \qquad (3)$$

$$y_i(n) = \varphi_j\,(v_j(n)) \qquad (4)$$

Where $y_i$ is the input from the previous layer $i$ which is then multiplied by the $j$th layer's weights w ji to produce $v_j(n)$, and finally $v_j(n)$ goes through a desirable activation ($\varphi_j$) to produce $y_i(n)$.

### B. *Long short-term memory (LSTM)*

LSTM is an extension of recurrent neural networks (RNN) [25]. These networks can transfer a concealed state as a representation of what has passed across the network. LSTMs have been successfully applied to a wide range of sequence classification issues [26]. LSTMs were developed in response to the problem of gradient loss in RNNs [27], which prevents the RNN's initial layers from being updated with a gradient vector. The difficulty is that as the local gradient reaches the first layers, it approaches 0 and the learning impact becomes negligible. With LSTM, a number of gates are created that can carry essential information for longer periods while also providing control over information flow. Each LSTM cell, like RNNs, sends a secret state to the next layer. The process for calculating the concealed state, on the other hand is different. This is accomplished by the employment of gates, which are discussed further below.

### Forget gate

This gate decides to maintain the preceding cell status information. The cell state is extremely important in conveying information across the LSTM cells. This technique can bestated as follows:

$$f_t = \sigma(w_f [h_{t-1} . x_t] + b_f) \qquad (5)$$

Where $x_t$ is the individual input, and $h_{t-1}$ is the last layer's hidden state. Then, using layer weights $w_f$ and a bias $b_f$, a linear transformation is performed. The output value ranges between 0 and 1, with a higher $f_t$ indicating that a greater fraction of the cell state is retained.

### Input gate

The data to be delivered to the current cell state is determined by the input gate. This technique is depicted in Equation 6:

$$i_t = \sigma(w_i[h_{t-1}. x_i] + b_i) \qquad (6)$$

It is the product of the prior hidden state $h_{t-1}$ and this cell's input $x_i$, which then undergoes a linear transformation with $w_i$, $b_i$, and finally a sigmoid activation $\sigma$ resulting in $i_t$. The calculations for the input gate and forget gate may appear to be the same, but the difference is in the amount of information flow that they can regulate. The forget gate modifies the cell state output, despite the fact that the input gate influences the information provided by the candidate gate, as will be detailed momentarily.

### Candidate gate

The candidate layer performs $tanh$ activation and is the primary source of the current cells' contribution to the cell state. This gate is expressed as follows:

$$c_t = tanh(w_c[h_{t-1}. x_c] + b_c) \qquad (7)$$

This gate is also a simple dot product of $h_{t-1}$ and $x_c$, which then undergoes the linear transformation with $w_c$ and $b_c$, each with its own weights and bias. Finally, the next cell state can be calculated as follows:

$$c_t = f_t \times c_{t-1} + i_t \times c_t \qquad (8)$$

### Output gate

This layer determines the following cell's concealed state as well as the current cell's final output. The expression can be shortened as follows:

$$o_t = \sigma(w_o[h_{t-1}. x_o] + b_o) \qquad (9)$$

This gate is also a simple dot product input $x_o$ and hidden state $h_{t-1}$, which is then linearly transformed with $w_o$ and $b_o$ before being transferred to a sigmoid function $\sigma$ to generate the cell output. The next hidden state, which will be passed to the next cell, will also be determined by performing a $tanh$ activation on the cell state $c_t$ and multiplying it by the cell output $o_t$:

$$h_t = o_t \times tanh(c_t) \qquad (10)$$

### C. Convolutional Neural network – Long Short-term memory (CNN – LSTM)

A recently released method known as CNN–LSTM, which was first applied on text classification [28], attempts to discover spatiotemporal relationships in data. The input in CNN–LSTM is first processed through a succession of convolutional layers to produce a sufficient feature map. These features will then be sent into a number of LSTM layers, which will inspect any potential temporal information. CNN–LSTM also use fully-connected layers for classification.

### D. Transformers

Transformers were first introduced for machine translation [22]. In the Time series, however, with contrast to text data, the signal is divided into $N$ segments of the same size, which is similar to the embedding operation on text features. Mathematically, the input data will be defined as:

$$X = x1, x2, ... X_N \qquad (11)$$

An example would be having a signal with a length of a thousand samples, where the feature vector could be a set of 4 time-series each having a sample size of $1000/4 = 250$. In transformers, in order to maintain the order of the series after separation, Positional Embeddings are used. In this layer an embedding is created based on the segment's max length and the total number of segments, which is then added to the feature vector. Keys, Values and Queries are also an important part of the transformer that help us calculate the attention weights, the formula for the aforementioned is as follows:

$$ATTENTION = softmax\left(\frac{QK^T}{\sqrt{d_k}}\right) V \qquad (12)$$

Where the attention weights are calculated using a softmax function, and $Q$, $K$, and $V$ stand for the queries, keys and values

that are basically dot products of the Input with their respective weights:

$$Q = XW_Q \qquad (13)$$

$$K = XW_K \qquad (14)$$

$$V = XW_V \qquad (15)$$

The last parameter $d_k$ is also known as the scaling factor. The Multi Head part of the transformer also points to the fact the attention result is consisted of different attention weights of different heads, also called heads that are combined together in the end to learn multiple representations of the data. It's also important to note that for a classification problem, only the decoder part of a transformer is needed and the encoder part is completely removed. The transformer architecture used in this paper is shown in Figure 2. The signals are converted to divergent sample intervals to model the importance of each interval.

*E. Evaluation*

The data is categorized into 70-30 format for training and testing. The reports are oriented towards the test results. This concludes 771 samples for training and 331 samples for testing for detecting depression. For Responder Group: 399 examples we used for training and 171 samples were left for testing. Different Versions of EEG segmenting were tested as transformer inputs here to find the optimal version. Divergent parameters were included for the transformer model in a way that the best accuracy with the minimal amount of model parameters is achieved.

## IV. Results

The software used for the analysis is Tensorflow. Table 1 shows the number of parameters in the tested models:

TABLE 1
PARAMETER COUNT FOR EACH MODEL

| Model | No. Params |
|---|---|
| Transformer | 34,169 |
| CNN1D | 34,855 |
| LSTM | 35,225 |
| CNN-LSTM | 33,789 |

Model sizes were chosen in a way that parameter numbers are really close. The number of heads was selected to be 4 in the transformer model. All the models were trained for 150 epochs and a batch size of 128. The early stopping condition was in a way that training would stop if the model was not improving on validation loss for more than 30 epochs. Adam optimizer was selected for faster training times and better performance with comparison to slower algorithms like SGD.

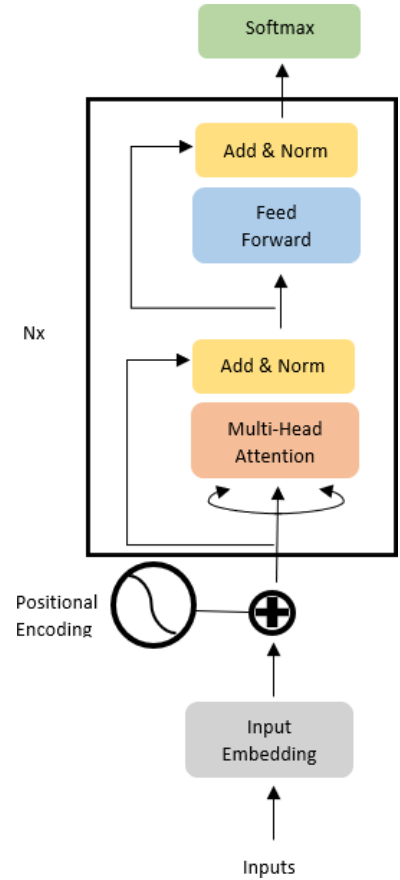

Fig. 3. Classification Procedure with transformer

Figure 3 shows the in-depth architecture of the transformer model adapted for this study:

Figure 4 demonstrates the architecture of the tested models:

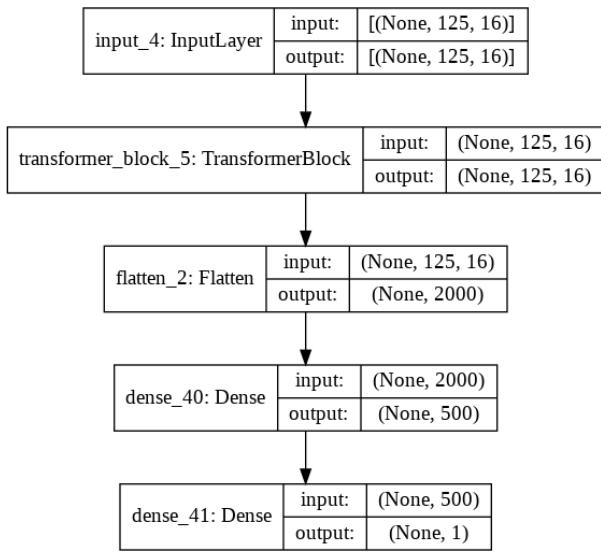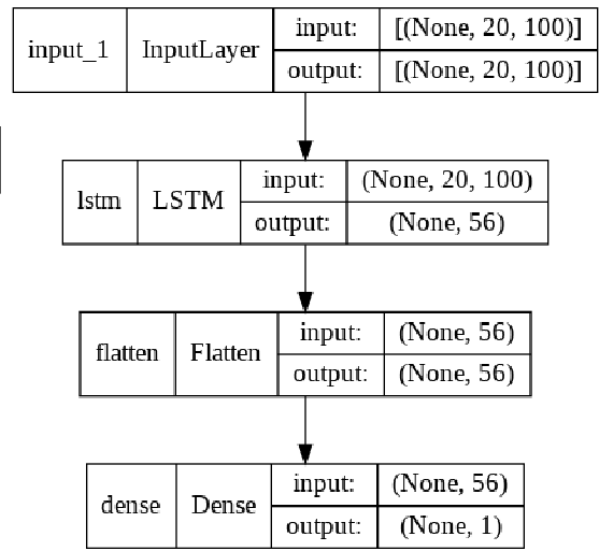

a) Transformer                     b) LSTM

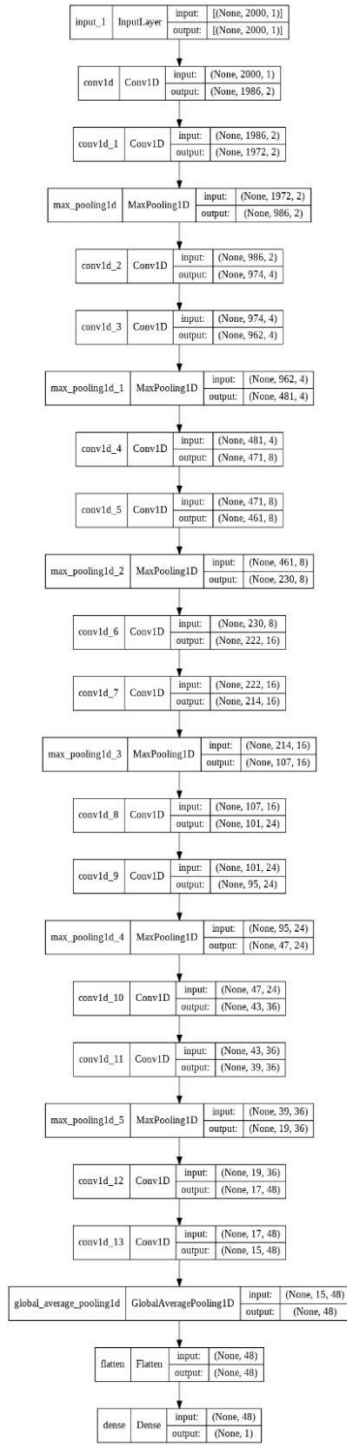

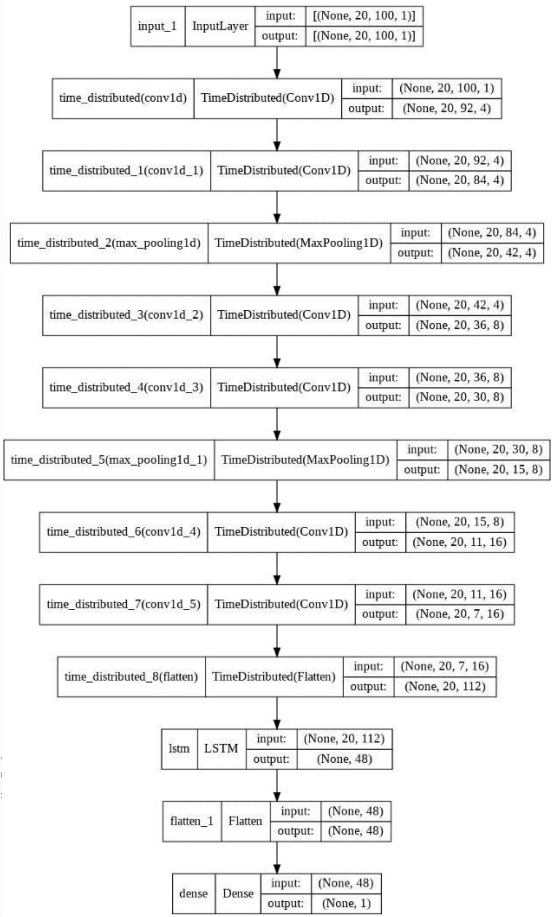

c) CNN 1D           d) CNN - LSTM

Fig. 4. Model Architectures

## A. Depression Classification

Figure 5 demonstrates the training and evaluation curves of the model:

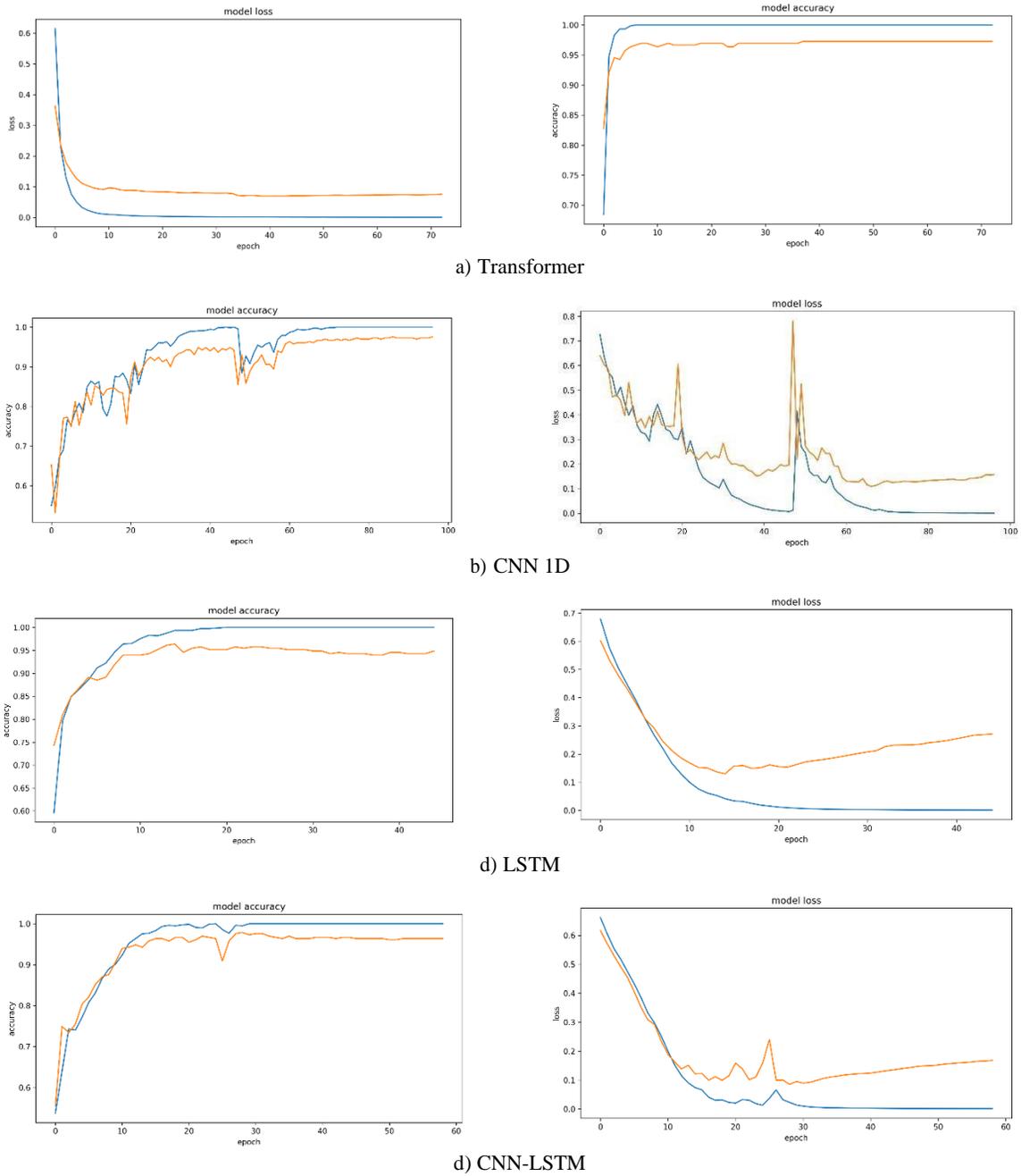

a) Transformer

b) CNN 1D

d) LSTM

d) CNN-LSTM

Fig. 5. Training and Evaluation Curves for the depression classification models for one training run

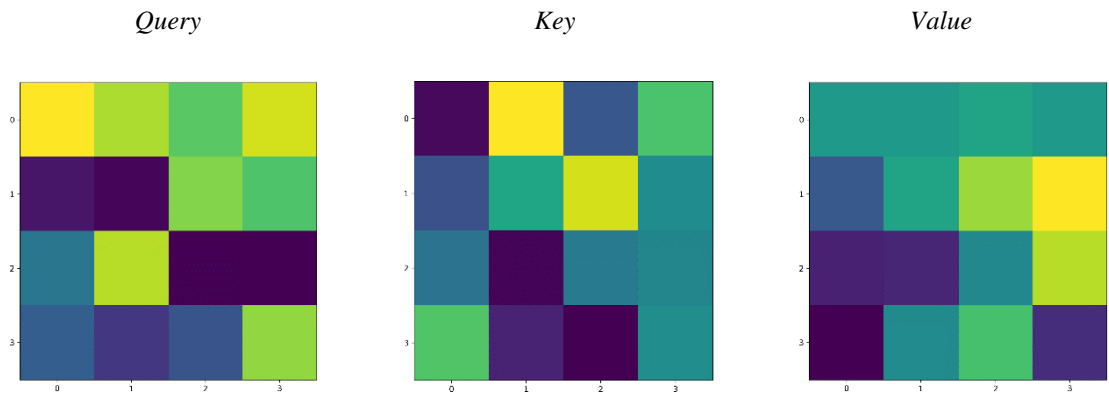

Fig. 6. Query, Key and Value Matrices of the Depression Classification Model

Table 2 visualizes the metrics acquired by the depression classification models over 10 initiation sessions

Figure 6 shows the color-coded Query, Key and Value matrices for the depression classification model:

TABLE 2
DEPRESSION CLASSIFICATION METRICS FOR THE CLASSIFACTION MODELS.

| Model | Precision | Accuracy | F1 | Recall | Specificity | AUC |
|---|---|---|---|---|---|---|
| Transformer | 95.42% ± 0.140 | 97.22% ± 0.007 | 97.37% ± 0.007 | 99.41% ± 0.003 | 94.87% ± 0.017 | 97.14% ± 0.008 |
| CNN1D | 96.29% ± 0.016 | 96.40% ± 0.131 | 96.52% ± 0.012 | 96.78% ± 0.159 | 96.00% ± 0.177 | 96.39% ± 0.131 |
| LSTM | 96.49% ± 0.009 | 95.04% ± 0.010 | 95.46% ± 0.010 | 94.48% ± 0.017 | 95.74% ± 0.012 | 95.11% ± 0.010 |
| CNN-LSTM | 97.26% ± 0.014 | 97.03% ± 0.009 | 97.13% ± 0.008 | 97.01% ± 0.010 | 97.06% ± 0.015 | 97.04% ± 0.009 |

Highest average Precision and Specificity was acquired by the CNN-LSTM model reaching up to 97.26% and 97.06%. Whereas the highest average Accuracy, F1, Recall and AUC belonged to the transformer model being 97.22%, 97.37%, 99.41%, 97.14% respectively. Following (Figure 7) is the confusion matrix of the models:

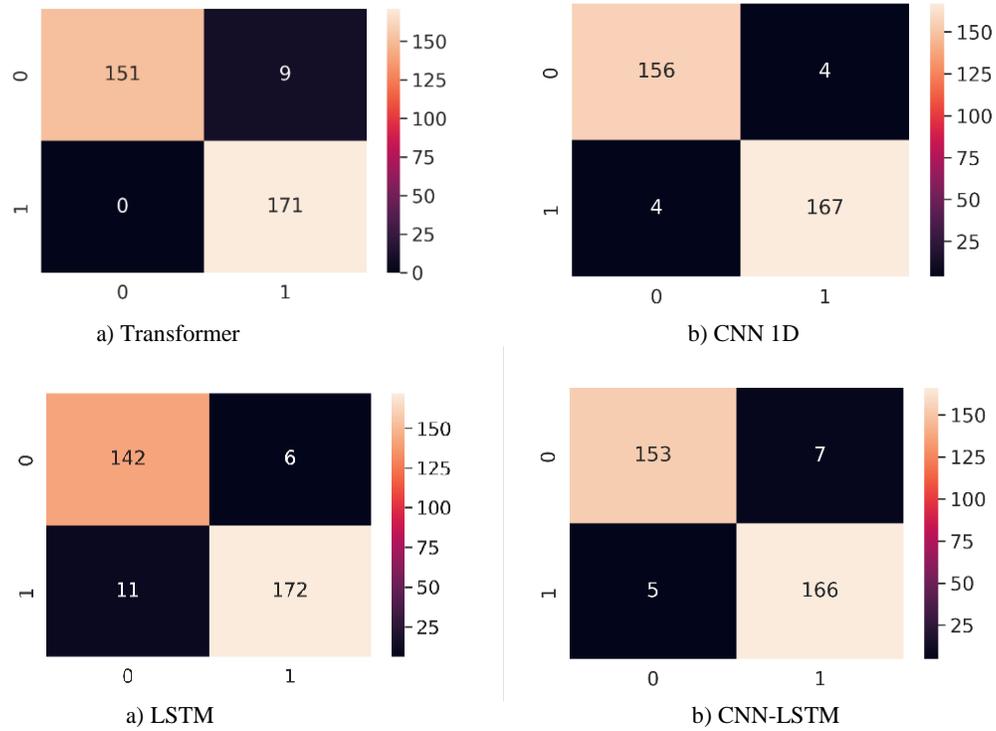

a) Transformer

b) CNN 1D

a) LSTM

b) CNN-LSTM

Fig. 7. Confusion Matrix for the depression classification models for one training run

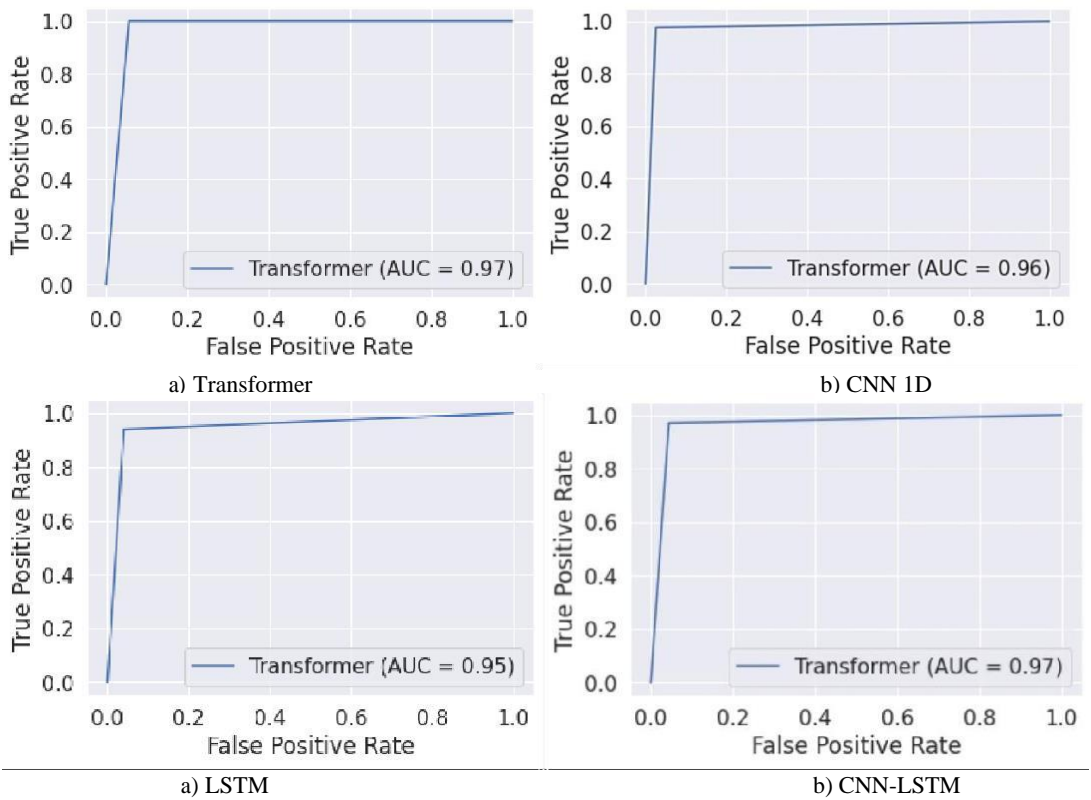

a) Transformer

b) CNN 1D

a) LSTM

b) CNN-LSTM

Fig. 8. AUC of the depression classification models

For the transformer on the best run, all 151 samples were correctly classified as normal, while 9 samples were mistakenly taken for depressed. For depressed samples, all the 171 were correctly detected. While no patient with depression was incorrectly classified as normal. Figure 8 is the Area Under Curve (AUC) of the depression classification models.

*B. Drug response classification*

Figure 9 demonstrates the training and evaluation curves of the responder classification models:

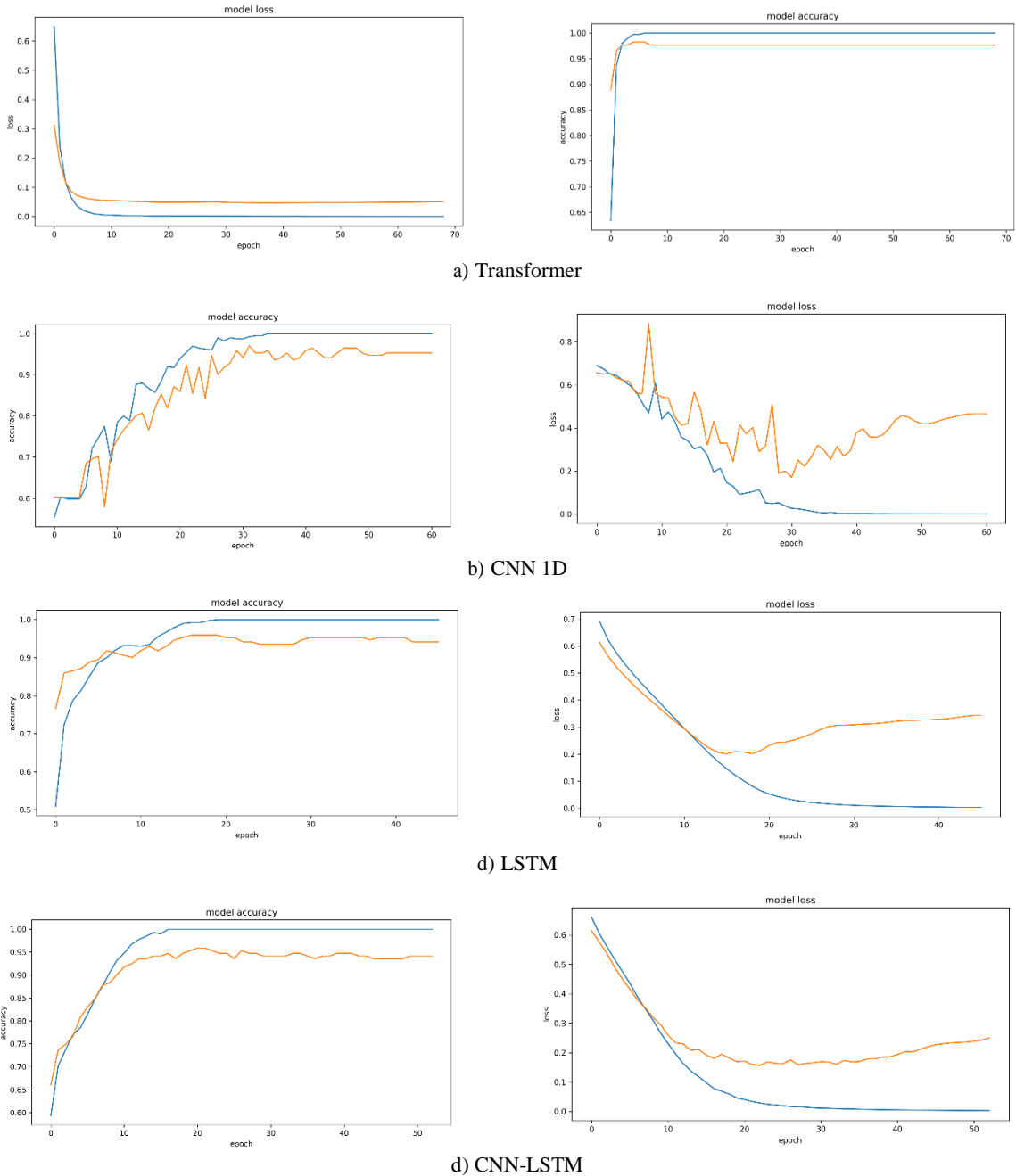

a) Transformer

b) CNN 1D

d) LSTM

d) CNN-LSTM

Fig. 9. Training and Evaluation Curves for the Responder Classification Models for one training run

Figure 10 shows the color-coded Query, Key and Value matrices for the responder classification model.

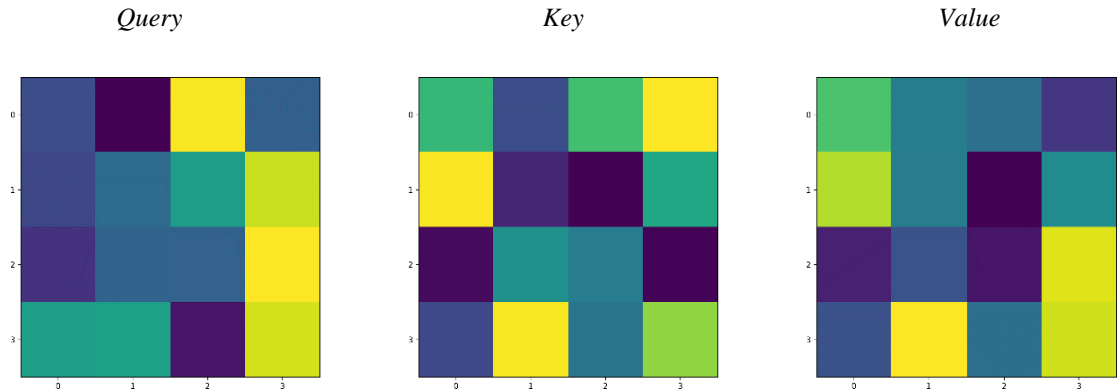

Fig. 10. Query, Key and Value Matrices of the Responder Classification Model

Figure 10 shows the color-coded Query, Key and Value matrices for the responder classification model.

Table 3 visualizes the metrics acquired by the responder classification models

Table 3 Responder Classification Metrics for the Transformer Model

| Model | Precision | Accuracy | F1 | Recall | Specificity | AUC |
|---|---|---|---|---|---|---|
| Transformer | 97.30% ± 0.005 | 97.01% ± 0.007 | 97.52% ± 0.006 | 97.76% ± 0.013 | 95.88% ± 0.017 | 96.82% ± 0.006 |
| CNN1D | 93.59% ± 0.021 | 93.56% ± 0.020 | 94.73% ± 0.016 | 95.92% ± 0.018 | 90.00% ± 0.034 | 92.96% ± 0.022 |
| LSTM | 93.20% ± 0.014 | 94.73% ± 0.017 | 95.75% ± 0.013 | 98.44% ± 0.015 | 89.11% ± 0.024 | 93.78% ± 0.018 |
| CNN-LSTM | 93.33% ± 0.012 | 94.44% ± 0.015 | 95.49% ± 0.012 | 97.76% ± 0.017 | 89.41% ± 0.020 | 93.58% ± 0.015 |

For responder classification, the transformer had the highest value for most of the experimented metrics: Precision, Accuracy, F1, Specificity and AUC. With Precision most of the models hovered around 93%. LSTM also acquired the highest f1 score below the transformer while having the best Recall 98.44% amongst all the models. In addition, low values of specificity were observed for CNN1D, LSTM and CNN-LSTM. Following (Figure 11) is the confusion matrix of the models. For the transformer, 66 samples were correctly classified as non-responders, whereas 2 samples were mistakenly taken for responders. For responder samples, 101 were correctly detected. While 2 patients that were not responders was incorrectly classified as responders.

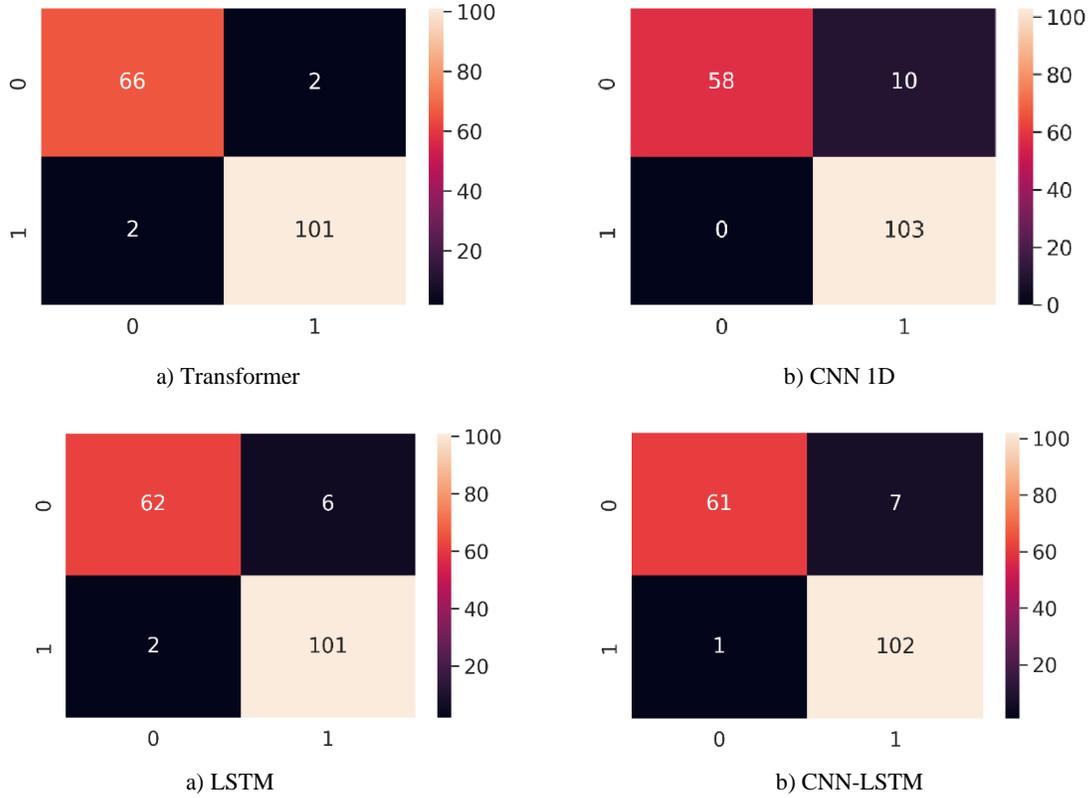

Fig. 11. Confusion Matrix for the Responder classification transformer models for one training run

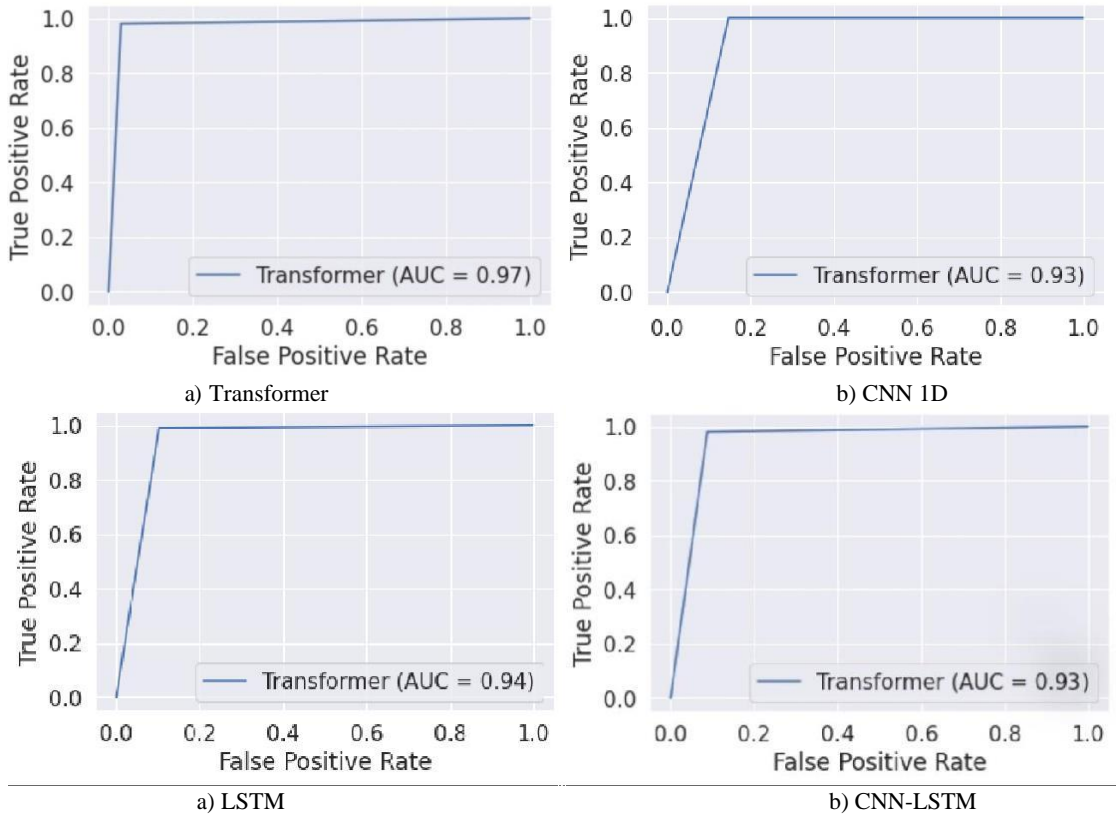

Fig. 12. AUC of the responder classification models

Figure 12 is the Area Under Curve (AUC) of the responder classification models.

## V. Discussion

In this work a framework based on state-of-the-art transformer networks was designed for automatic detection of MDD patients, as well as their reaction to medications. In comparison to other Deep Learning Methods such as CNNs, LSTMs, and combinations of them, the model achieved 99.41% sensitivity while having a relatively low number of parameters. According to table 4, the transformer model outperformed other strategies in terms of computing efficiency and accuracy. This can be attributed to the use of attention mechanisms and positional embeddings. In addition, based on the loss plots in figure 9, it can be concluded that the transformer is far less prone to overfitting and has a more stable training process compared to its counterparts, namely CNN, LSTM and CNN-LSTM. The results of this study are compared in to new related work that used EEG signals from the identical source [23, 19] and a different source [11, 17, 15] in table 4. As it can be inferred, the accuracy achieved in this study is comparable to that attained in previous studies using deep learning methods, and this study acquires the highest rate of sensitivity with respect to recent work, demonstrating the viability of the proposed technique.

TABLE 4 COMPARISON OF RECENT STUDIES

| Authors | Methods | Samples | Classification Methods | Model Parameters | Sensitivity | Specificity | Accuracy |
|---|---|---|---|---|---|---|---|
| Acharya et al [11] (2018) | 13-layer CNN | 15 Depressed 15 normal | 1D CNN | Not Reported | 94.99% | 96.00% | 95.96% |
| Ay et al [17] (2019) | - | 15 Depressed 15 normal | CNN-LSTM | Not Reported | 98.55 % | 99.70% | 99.12%(Right) 97.66%(Left) |
| Mumtaz and Qayyum (2019) [23] | - | 33 Depressed 30 normal | CNN-LSTM, 1D-CNN | Not Reported | 98.34% | 99.78% | 98.20% |
| Saeedi (2020) [19] | Effective Connectivity (GPDC, dDTF) | 33 Depressed 30 normal | IDCNN, 2DCNN, 1DCNN-LSTM, 2DCNN-LSTM | | 98.52% | 100% | 99.25% |
| Wan et a l[15] (2020) | - | 12 patients without a prescription, 12 drugs, 11 normal | Hybrid EEGNet | Not Reported | 68.78% | 84.45% | 79.08% |
| Current Study | - | 33 Depressed 30 normal | Transformer CNN1D LSTM CNN-LSTM | ~ 30k | 99.41% 96.78% 94.48% 97.01% | 94.87% 96.00% 95.74% 97.06% | 97.22% 96.40% 95.04% 97.03% |

It's also important to note that this work has the advantage of using Transformers for sequence learning compared to other deep learning models, also current sequence-based models were studied thoroughly. We suggest that other models cannot be as competent when having low number of parameters, although with hundreds of thousands of them good results can be achieved even on CNN and LSTM models and their variants, but this, of course not suitable for real life scenarios where every second of the patient's time and the hardware usage can have heavy impacts on the feasibility. The main drawback of the research can be considered the dataset size. By performing regularization terms and simplifying deep models we were able to overcome the problem.

## VI. Conclusion

Signals from 34 depressed individuals and 30 healthy subjects were analyzed in this study utilizing transformers to identify depression and discriminate between depressed people who responded to medication and those who did not. The method presented in this paper has the advantage of having a low computational load and a high speed which achieved 97.14% accuracy in distinguishing between depressed and healthy samples and 97.01% for detecting responder participants. Based on the results, the recommended tool has the ability to help clinicians examine MDD patients for earlier diagnosis and management of their situation as a monitoring instrument.

## VII. COMPLIANCE WITH ETHICAL STANDARDS

Conflict of interest: The authors declare that they have no conflict of interest. Ethical approval: The dataset was approved by the ethics committee in Hospital University Sains Malaysia (HUSM), Kelantan, Malaysia.